\documentclass[a4paper,11pt]{article}
\pdfoutput=1 % if your are submitting a pdflatex (i.e. if you have
\usepackage{jcappub} % for details on the use of the package, please
\usepackage[T1]{fontenc} % if needed
\usepackage{float}
\usepackage{hyperref}

\title{Study of Gravastars in Rastall Gravity}
\author[a]{Shounak Ghosh,}
\author[b]{Sagar Dey \note{Corresponding author}}
\author[a]{Amit Das,}
\author[b]{Anirban Chanda,}
\author[c,1]{and B C Paul.}

\affiliation[a]{Department of Physics, Indian Institute of Engineering Science and Technology,\\Shibpur, Howrah 711103, West Bengal, India}
\affiliation[b]{Department of Physics, University of North Bengal, West Bengal 734014, India.}
\affiliation[c]{IUCAA Centre for Astronomy Research and Development (ICARD) and Department of Physics, University of North Bengal, West Bengal 734014, India.}

\emailAdd{shounakphysics@gmail.com}
\emailAdd{sagardey231@gmail.com}
\emailAdd{amdphy@gmail.com}
\emailAdd{anirbanchanda93@gmail.com}
\emailAdd{bcpaul@associates.iucaa.in}

\abstract{Gravastars have been considered as a feasible alternative to black holes in the past couple of decades.
Stable models of gravastar have been studied in many of the alternative gravity theories besides standard
General Relativity (GR). The Rastall theory of gravity is a popular alternative to GR, specially
in the cosmological and astrophysical context. Here, we propose a stellar model under the Rastall gravity
following Mazur-Mottola's \cite{Mazur2001,Mazur2004} conjecture. The gravastar consists of three regions, viz., ($I$) Interior region,
($II$) Intermediate shell region, and ($III$) Exterior region. The pressure within the interior core region
is assumed with a constant negative matter-energy density which provides a  repulsive
force over the entire thin shell region. The shell is assumed to be made up of fluid of ultrarelativistic plasma
which follows the Zel'dovich's conjecture of stiff fluid \cite{Zeldo1962,Zel'dovich1972}. It is also assumed that the pressure is
 proportional to the matter-energy density according to Zel'dovich's conjecture, which cancel the
repulsive force exerted by the interior region. The exterior region is completely vacuum  which is
described by the Schwarzschild-de Sitter solution. Under all these specifications we obtain a set of exact and singularity-free solutions of the gravastar model presenting several physically valid features within
the framework of Rastall gravity. The  physical properties of the shell region namely, the energy density, proper length, total energy and entropy are explored. The  stability of the gravastar model is investigated using   the surface redshift  against the shell thickness and  maximizing the entropy  of the shell  within the framework  of Rastall gravity.}


\begin{document}
\maketitle
\flushbottom

\section{Introduction}\label{sec1}
The end point of the stellar evolution of a massive object always attract the attention of the astrophysicists.
It is  conjectured that in the final stage of stellar evolution if the mass of the stellar
remnant after forming planetary nebula or initiating supernova explosion is greater than three solar
mass, then the collapsing phase of the star due its self gravitational pull leads to a highly dense
object, {\it i.e.}, black hole. In 1916, Schwarzschild obtained the simplest black hole solution
from the Einstein field equation in vacuum which is considered as the important solution of a classical
black hole (static as well as uncharged). But the Schwarzschild  solution has some shortcomings due to
(i) presence of singularity inside the black hole, (ii) existence of event horizon which leads to
major unsolved issues. Mazur and Mottola~\cite{Mazur2001,Mazur2004} first  proposed a
new idea of collapsing stellar object by considering an extended idea of Bose Einstein Condensation
(BEC) in the gravitating system. In the model it was proposed that  one can construct  a cold, dark and compact object, they  nomenclature it as a Gravitational Vacuum Condensed Star or Gravastar. The model of gravastar satisfies all the
theoretical criterion for a stable end point of the stellar evolution and provides solution to  the
problems of classical black hole.
After the successful detection of gravitation wave (GW) in 2015 \cite{Abbott2016}, it is assumed that  the
GWs are originated due to merging of two massive black holes. As the observational signals does not confirm the recovery of the basic problems of black holes, it demands an alternative approach to describe the shortcomings.
 In this context, the gravastar may play an important role to describe the final state  of the
stellar evolution. Though there are no sufficient observational evidences in favour of  gravastars
 directly for their existence, it is important to study the  concept of gravastar that can be
claimed as a viable alternative  to resolve the conceptual issues in understanding the black holes.\\

The gravastar consists of three regions namely interior region,  intermediate thin shell, and
exterior region. As proposed by  Mazur  and Mottola \cite{Mazur2001,Mazur2004} the interior region
of the gravastar being in de Sitter condensate phase is assumed to be filled with vacuum energy,
the exterior region is completely vacuum known as Schwarzschild vacuum, and those two regions are
separated by a thin shell of ultra-relativistic matter of very high density.  The barotropic  equation of
state (EOS)  is   $p=\omega \rho$, where
$p$ and $\rho$ are the pressure and matter-energy density respectively and $\omega$ is the EOS parameter.

In a gravastar  the EOS parameter $\omega$, is different in different regions as follows:\\
\noindent ($I$) Interior Region : $0 \leq r < R_1$, $\omega=-1$, $i.e.$, EOS is $p =- \rho$ ,\\
($II$) Thin Shell  Region: $ R_1 <r < R_2$, $\omega=1$, $i.e.$, EOS is $ p = \rho$ ,\\
($III$) Exterior Region : $R_2 < r $, $\omega=0$, $i.e.$, EOS is $ p =\rho= 0$ .\\
where $R_1$ and $R_2$ represent the radii of the interior and exterior boundary of the
gravastar respectively. Therefore, thickness of the intermediate shell  region is given by
$R_2-R_1 =\epsilon\ (say)$, where $\epsilon\ll 1$ because in  a gravastar the thickness of the shell is very small compared to its size.

The idea of gravastar is discussed  in the literature \cite{Visser2004,Cattoen2005,Carter2005,Bilic2006,Lobo2006,DeBenedictis2006,Lobo2007,Horvat2007,Cecilia2007,
Rocha2008,Horvat2008,Usmani2008,Nandi2009,Turimov2009,Harko2009,Usmani2011,Rahaman2012a,Rahaman2012b,Bhar2014,
Rahaman2015,Ghosh2017,Ghosh2018,Ghosh2019a,Ghosh2019b,chan1,chan2} as an alternative to BH.
Mazur and Mottola proposed a three layer structure of the gravastar and studied the
thermodynamical stability employing the entropy maximization technique, whereas Visser~\cite{Visser2004} showed the dynamical stability of the system.
Based on the three layer system it is now
established that the theoretical acceptance of gravastar model in both lower and higher dimensions with or
without charge is important in astrophysics. Usmani et al. \cite{Usmani2008}
shown the possibility of existence of higher dimensional gravastar, and later on Bhar \cite{Bhar2014}
and Ghosh et al.~\cite{Ghosh2017}  studied the higher dimensional gravastar with and without conformal motion respectively. The possibility of gravastar  in lower dimension is also studied \cite{Ghosh2019a,Ghosh2019b}. Chan and Silva \cite{chan1}  have investigated the effect of  charge on the stability of the gravastar formation considering a charged shell configuration. Again, considering  the Vaidya exterior spacetime, Chan $et ~al.$ \cite{chan2} have studied the dynamical models of prototype radiating gravastars and showed that the final output may corresponds to a number of objects, a black hole, an unstable gravastar, a stable gravastar or a ``bounded excursion” gravastar depending on the mass of the shell that evolves in time, the cosmological constant, and the initial position of the dynamical shell. All of these works are mainly carried out in the framework of Einstein's general theory of relativity (GR). Though it is well known that Einstein's GR is one of the most promising theory for revealing most of the
hidden mysteries of the Nature it is realized that a modification of the GR is inevitable to wipe out
shortcomings in the theoretical as well as observational aspects. The confirmation
of accelerated universe based on some observational evidences as well as the existence of dark matter and dark
energy led to a theoretical challenge to GR ~\cite{Riess1998,Perlmutter1999,Bernardis2000,Hanany2000,Peebles2003,Padmanabhan2003,Clifton2012,Riess2007,Tegmark2004,Amanullah2010,Komatsu2011}. To accommodate the present acceleration of the universe a number of alternative theories of gravity, namely $f(R)$ gravity, $f(\mathbb{T})$ gravity, $f(R,\mathcal{T})$ gravity, $f(\mathbb{T},\mathcal{T}) $ gravity are developed successively with a modification either in the geometrical part or in the matter-energy part of Einstein field equations. In most of these theories different functions of the Ricci scalar with or without other scalar quantities
are incorporated in the gravitational Lagrangian of the  corresponding Einstein-Hilbert action. A lot of works have been performed on compact stars as well as gravastars under the framework of modified theories of gravity as available in the literature~\cite{Das2015,Das2016,Das2017,Biswas2019,Ghosh2020,Das2020,Sengupta2020,Banerjee2020}. But, in most these theories the energy-momentum tensor (EMT) is not generally conserved unlike the case of general relativity. This interesting aspect of modified gravity  motivates us to make a general study of gravastar under the Rastall theory of gravity as proposed by Rastall
in 1972~\cite{Rastall1972, Rastall1976} based on the idea that the usual law of conservation of energy-momentum tensor, i.e., $\nabla_\nu T_\mu^\nu=0$ is violated when a curved spacetime is taken into account. The violation
of conservation law of EMT is the key point of this modification where the covariant derivative of
EMT is directly proportional to the divergence of scalar curvature. Mathematically, it can be written as
$\nabla_\nu T_\mu^\nu \propto \mathcal{R}_{,\;\mu}$, which further takes the form $\nabla_\nu T_\mu^\nu =\psi \mathcal{R}_{,\;\mu}$, where
$\psi$ is the constant of proportionality. \\

Various cosmological as well as astrophysical
aspects have been investigated under the framework of Rastall theory due  its numerous appealing features and simpler form of field equations as available in the literatures~\cite{Capone2010,Batista2012,Heydarzade2017, Haghani2018,Heydarzade2018,Kumar2018,KNABH,Ma2017,Oliveira2015,Bhar2020}. Recently, in astrophysical context Rastall gravity has been extensively investigated in the context of rotating black holes~\cite{Kumar2018},
Kerr-Newman-AdS black hole \cite{KNABH}, non commutative geometry~\cite{Ma2017}. Rastall theory is also successful in the context of relativistic solutions of compact stars. Oliveira $et ~al.$~\cite{Oliveira2015} have studied Neutron stars in Rastall gravity.  Bhar et al.~\cite{Bhar2020}  studied a highly dense compact object in  Rastall gravity
using Tolman-Kuchowicz spacetime. \\

The outline of the present investigation is as follows: In section \ref{sec2} the basic mathematical formalism
of Rastall theory is presented and the explicit form of the field equations
along with the conservation equation of Rastall gravity is given,   in
section \ref{sec3}, we obtain the solutions of the field equations considering the different regions, viz., interior
region, exterior region, and  shell region of the gravastar. In section~\ref{sec4}, we consider  the matching
conditions in order to determine the numerical values of several constants which arise in different
calculations. Section~\ref{sec5} deals with some physical features of the model, {\it i.e.}, proper length, pressure, energy, and entropy within the shell region. Junction conditions are presented in section~\ref{sec6}. Stability issue of the model has been explored in  section~\ref{sec7}, and finally, we pass
some concluding remarks in section~\ref{sec8}.

\section{The Field Equations geometry in Rastall theory}\label{sec2}

We start from the short introduction to Rastall theory of gravity which was introduced by P. Rastall \cite{Rastall1972, Rastall1976}. According to the original idea of Rastall \cite{Rastall1972}, the vanishing of covariant divergence of the matter energy-momentum tensor is no longer valid and this vector field is proportional to the covariant derivative of the Ricci curvature scalar as,
\begin{equation}
\label{eq1}
    \nabla_{\nu} T_{\mu}^{\nu}= a_{\mu}
\end{equation}
where the functions $a_{\mu}$ vanish in flat space-time, but not in curved space-time, and which are in agreement with present observations. Here, it is noted that $a_{\mu} = \psi \mathcal{R}_{,\;\mu}$,  where $\psi$ is a constant, $\mathcal{R} =g^{\mu\nu}\mathcal{R}_{\mu\nu}= g^{\mu\nu}g^{\lambda\pi}\mathcal{R}_{\lambda\mu\nu\pi}$ is the curvature invariant, $\mathcal{R}$ and $T_{\mu}^{\nu}$ represent the Ricci scalar and energy momentum tensor respectively and the comma denotes the partial derivative. The primary mathematical relation of Rastall gravity in terms of energy-momentum tensor and curvature scalar is defined by,
\begin{equation}
\label{eq2}
\mathcal{R}_{\mu\nu} + (k\psi-\frac{1}{2})g_{\mu\nu} \mathcal{R} = k T_{\mu\nu}
\end{equation}
The Eq.~(\ref{eq2}) leads to a modified theory whose field equations can be
written as
\begin{equation}\label{eq3}
\mathcal{G}_{\mu\nu} + k \; \psi g_{\mu\nu}\mathcal{R} = k T_{\mu\nu}
\end{equation}
where $\mathcal{G}_{\mu\nu} $ represents Einstein tensor. In the above set of field equations, the parameter $k$ plays the role of gravitational coupling parameter for modified Rastall theory. Contracting Eq.~(\ref{eq2}) one
 obtains $\mathcal{R} (4k\psi- 1)= k T^{\mu}_{\;\mu}$. We exclude the case $ k \psi = \frac{1}{4}$, since it implies that $T^{\mu}_{\;\mu}=0 $, which is not always true. In empty space-time one has $T_{\mu \nu}=0$, and hence  $\mathcal{R}=0$. The field equations then reduce to $\mathcal{R}_{\mu\nu}=0$, just as in the Einstein theory. In this study, we define $\xi = k \psi$ as the Rastall parameter, which can be considered as dimensionless. Consequently, the gravitational coupling parameter $k$ and the constant $\psi$ must satisfy:
\begin{equation}\label{eq4}
 k= \frac{8 \pi G}{c^{4}} \left (\frac{4\xi-1}{6\xi-1} \right)
\end{equation}
where $G$ is the Newtonian gravitational constant, and $c$ is the speed of light. When $\xi= 0$, $k$ becomes the Einstein gravitational constant ($k= \frac{8 \pi G}{c^{4}}$). It is also noted that $\xi= \frac{1}{4}$ is not permitted here. Thus, a new set of field equations for Rastall theory is given by
\begin{equation}\label{eq5}
    \mathcal{G}_{\mu\nu} + \xi\; g_{\mu\nu}\mathcal{R} = 8 \pi T_{\mu\nu} \left[\frac{4\xi-1}{6\xi-1} \right],
\end{equation}
To study the gravastar in $(3 + 1)$ dimension we first consider the line element for the interior spacetime of
a static spherically symmetric matter distribution is in the form
\begin{equation}\label{eq6}
 ds^{2}= - e^{\nu(r)} dt^{2}+ e^{\mu(r)} dr^{2}+ r^{2}(d\theta^{2}+sin^{2}\theta d\phi^{2}),
\end{equation}
where $\mu$ and $\nu$ are the metric potentials which are function of r only.

Now we assume that the matter distribution in the interior of the star is that of a perfect fluid type and can be given by
\begin{equation}\label{eq7}
    T_{\mu\nu}= (\rho+ p)u_{\mu}u_{\nu} + p g_{\mu\nu},
\end{equation}
where $\rho$ represents energy density, $p$ is the isotropic pressure and $u^{\mu}$ represents four-velocity of
the fluid under consideration such that $u^{\mu}= e^{-\nu/2} \delta^{\mu}_{0}$ with $u_{\mu}u^{\mu}= - 1$.

Thus in Rastall gravity Eq. (\ref{eq5}) along with Eq.~(\ref{eq6}) and Eq.~(\ref{eq7}) yields the following
set of equations:
\small
\begin{eqnarray}\label{eq8}
\frac{\left(4\xi-1\right)\rho(r)}{6\xi-1}= e^{-\mu}\left[\frac{\mu'}{r}-\frac{1}{r^2}\right] +\frac{1}{r^2}+\xi e^{-\mu}
\left[\nu'' +(\nu')^2- \mu'\nu' -\frac{2}{r}(\mu' -\nu') -\frac{2\left(e^{\mu}-1\right)}{r^2}\right],
\end{eqnarray}
\normalsize

\small
\begin{eqnarray}\label{eq9}
\frac {(4\xi-1) p(r)}{6\xi-1}=e^{-\mu}\left[\frac{\nu'}{r}+\frac{1}{r^{2}} \right] -\frac{1}{r^{2}}-e^{-\mu}\xi
\left[\nu'' + (\nu')^2- \mu'\nu' -\frac{2}{r}(\mu' -\nu') -\frac{2\left(e^{\mu}-1\right)}{r^2}\right],
\end{eqnarray}
\normalsize
\[
\frac{(4\xi-1)p(r)}{6\xi-1}=\frac{e^{-\mu}}{4}\left[ 2\nu''+\left(\nu'\right)^2+ \frac{2\nu'}{r}- \frac{2\mu'}{r}- \mu'\nu'\right]
\]
\begin{eqnarray}\label{eq10}
- e^{-\mu}\xi\left[\nu'' + (\nu')^2- \mu'\nu' -\frac{2}{r}(\mu' -\nu') -\frac{2\left(e^{\mu}-1\right)}{r^2}\right].
\end{eqnarray}
 Here the symbol `$\prime$' denotes differentiation with
respect to the radial parameter $r$ and the geometrized units $G=c=1$.

Therefore, the energy conservation equation in Rastall gravity can be written as
\begin{eqnarray}\label{eq11}
p' +\frac{1}{2}(p+\rho) \nu'-\frac{\xi}{{4\xi-1}} \left(\rho' -3p'\right)=0.
\end{eqnarray}
The above equation is different from that obtained in GR and can be retrieved in the limit $\xi \rightarrow 0$.

\section{The gravastar models}\label{sec3}
Now we will study the three regions of gravastar precisely in Rastall theory of gravity.

\subsection{Interior spacetime}
According to Mazur-Mottola \cite{Mazur2001,Mazur2004}, the interior region
of gravastar can be described by the EOS as
\begin{equation}\label{eq12}
    p = - \rho
\end{equation}
This negative (repulsive) pressure acting radially outward from the centre of the spherically
symmetric gravitating system to counter balance the inward gravitational pull of the shell. The above EOS is known in
the literature as a `degenerate vacuum'or`$\rho$-vacuum'~\cite{Davies1984,Blome1984,Hogan1984,Kaiser1984}
which is a gravitational BEC after the phase transition occurring at the horizon (replaced
by a shell for a gravastar) and it acts along the radially outward direction to oppose the
collapse to be continued. Plugging this EOS in the conservation equation as provided by Eq.~(\ref{eq12}), we get that
\begin{equation}\label{eq13}
p=-\rho=-\rho_{c}.
\end{equation}
Here $\rho_{c}$ is the constant interior density, thus implying constant pressure.

Using Eq. (\ref{eq13}) in the field Eqs.~(\ref{eq8}) and (\ref{eq9}), we get the expression of the
metric potential $\mu$ given by
\begin{equation}\label{eq14}
\rm e^{-\mu(r)}=\frac{1}{3}C_1 r^2-\frac{C'}{3r}+1.
\end{equation}
where $C_1$ and $C'$ are integration constants. To make the solution  regular at $r=0$ one can
demand that $C'=0$. Thus, Eq.~(\ref{eq14}) essentially reduces to
\begin{equation}\label{eq15}
 \rm e^{-\mu(r)}=\frac{1}{3}C_1 r^2+1.
\end{equation}

Again, from Eqs.~(\ref{eq9}) and (\ref{eq10}) in the interior the another metric potential can be obtained as,
\begin{equation}\label{eq16}
 \rm e^{-\nu(r)}=C_2\left(\frac{1}{3}C_1 r^2+1\right).
\end{equation}

Here $C_2$ is an constant of integration. From the above solutions it can be noticed that the interior
solutions have no singularity and thus the problem of the central singularity of a classical black hole can be averted.

The active gravitational mass of the interior region can be found by using the following equation as
\begin{equation}\label{eq17}
\tilde{M}=\int^{R_1=R}_0 4\pi r^2 \rho dr=\frac{4\pi R^3 \rho_c}{3}.
\end{equation}

\subsection{Intermediate thin shell}
The shell consists of ultra relativistic fluid and it obeys the EOS $ p= \rho$. In connection to cold
baryonic universe, Zel'dovich~\cite{Zeldo1962,Zel'dovich1972} first conceived the idea of this kind of fluid and
it is also known as the stiff fluid. In the present case we can argue that this may arise from the
thermal excitations with negligible chemical potential or from conserved number density of the gravitational
quanta at the zero temperature. Several authors have extensively exploited this type of fluid to explore
various cosmological~\cite{Madsenet1992,Carr1975,Chakraborty2001} as well as astrophysical phenomena~\cite{Linares2004,Braje2002,Buchert2001}.
Within the non-vacuum region, i.e., the shell one
can observe that it is very difficult to find solution of the field equations. However, within the framework
of the thin shell limit, i.e., $ 0 < e^{-\mu(r)} \ll 1 $, it is possible to find an analytical solution.
As prescribed by Israel we can possibly argue that the intermediate region between the two spacetimes
must be a thin shell. Also within the thin shell region any parameter which is a function of r is, in
general, $ \ll 1 $ as $ r \rightarrow 0$. Due to this kind of approximation along with the above EOS as
well as Eqs.~(\ref{eq8}), (\ref{eq9}) and (\ref{eq10}), one can obtain the following equations:

\begin{equation}\label{eq18}
\frac{(4\xi-1)\rho(r)}{(6\xi-1)}={\frac{{e^{-\mu }}\mu' }{r}} + \frac{1}{r^2}+{e^{-\mu }}\xi\left( -\mu' \nu' -\frac{2\mu' }{r} \right) -\frac{2\xi}{r^2},
\end{equation}

\begin{equation}\label{eq19}
\frac{(4\xi-1)p(r)}{(6\xi-1)}=-\frac{1}{r^2}-{e^{-\mu}}\xi\left(-\mu' \nu' -\frac {2\mu'}{r}\right)+\frac{2\xi}{r^2},
\end{equation}

\begin{equation}\label{eq20}
\frac{(4\xi-1)p(r)}{(6\xi-1)}=\frac{e^{-\mu}}{4}\left(-\mu'\nu' -\frac{2\mu'}{r}\right) -{e^{-\mu}}\xi\left(- \mu' \nu' -{\frac{2\mu'}{r}}\right)+\frac{2\xi}{r^2}.
\end{equation}

From Eqs.~(\ref{eq19})-(\ref{eq20}) we get
\begin{equation}\label{eq21}
\nu' =\frac {-2e^{-\mu} \mu' r+4}{e^{-\mu} r^2{\mu' }}.
\end{equation}

Now using Eqs.~(\ref{eq18}) and (\ref{eq19}) along with the above condition we have obtained the two
metric functions of the shell as follows,

\begin{equation}\label{eq22}
 e^{-\mu(r)}= -12\xi \ln(r)+2 \ln(r) + C_{3},
\end{equation}

\begin{equation}\label{eq23}
\nu= \frac{4}{(6\xi-1)} \ln(r) - \frac{12\xi}{(6\xi-1)} \ln(r) + C_{4}.
\end{equation}
where $C_{3}$ and $C_{4}$ are another two integration constants.

\subsection{Exterior spacetime}
The exterior of the gravastar is assumed to obey the EOS $p=\rho=0$ which means that the outside region
of the shell is completely vacuum. Thus, using Eqs.~(\ref{eq8}) and (\ref{eq9}), we get
\begin{equation}\label{eq24}
    \mu' + \nu' = 0
\end{equation}
Considering the solution of Eq.~(\ref{eq24}), one can consider the line element
for the exterior region as the well-known Schwarzschild de Sitter metric which is given by
\begin{equation}\label{eq25}
ds^2=-\left(1-\frac{2M}{r}-\frac{\Lambda r^2}{3}\right)dt^2+\left(1-\frac{2M}{r}-\frac{\Lambda r^2}{3}\right)^{-1}dr^2+r^2(d\theta^2+\sin^2\theta d\phi^2),
\end{equation}
where $M$ is the total mass and $\Lambda$ is the cosmological constant.

\section{Boundary Condition}\label{sec4}
The gravastar configuration has two boundaries, one is between interior region and intermediate thin shell (i.e., at $r=R_1$)
and the other is between the shell and exterior spacetime (i.e., at $r=R_2$). For any stable configuration the metric
functions must be continuous at these interfaces. In order to determine the unknown constants of our present study viz.
$C_{1}$, $C_{2}$, $C_{3}$, $C_{4}$ and $\Lambda$, we have matched the metric functions at these
boundaries and eventually obtained the values of these constants:

(i) At the boundary between the interior and the intermediate thin shell, $i.e.$ at $R= R_{1}$,
\begin{eqnarray}
\frac{1}{3} C_1 R_1^2+1=C_3-12 \xi  \log \left(R_1\right)+2 \log \left(R_1\right),\label{eq26}\\
C_2 \left(\frac{1}{3} C_1 R_1^2+1\right)=C_4 R_1^{\frac{4-12 \xi }{6 \xi -1}}.\label{eq27}
\end{eqnarray}

(ii) Again, from the continuity of the metric potentials and $\frac{\delta g_{tt}}{dr}$ at $R= R_{2}$ we get another three conditions
\begin{eqnarray} \label{eq27}
-\frac{2 M}{R_2}-\frac{1}{3} \Lambda  R_2^2+1=C_3-12 \xi  \log \left(R_2\right)+2 \log \left(R_2\right),\label{eq28}\\
-\frac{2 M}{R_2}-\frac{1}{3} \Lambda  R_2^2+1=C_4 R_2^{\frac{4-12 \xi }{6 \xi -1}},\label{eq29}\\
\frac{2 M}{R_2^2}-\frac{2 \Lambda  R_2}{3}=\frac{C_4 (4-12 \xi ) R_2^{\frac{4-12 \xi }{6 \xi -1}-1}}{6 \xi -1}.\label{eq30}
\end{eqnarray}

To obtain the numerical values of five different constants, i.e.,  $C_{1}$, $C_{2}$, $C_{3}$, $C_{4}$
and $\Lambda$ we choose the mass of the gravastar $M= 3.75 M_{\odot}$, interior radius $R_{1} = 10$ km and
the four different values of $\xi = 0.228, 0.23, 0.235$ and $0.24$ which are enlisted in Table-I. We have
also shown the predicted values of $R_{2}$ for different $\xi$ values from the concept of the surface energy
density ($\Sigma$) (from section 6) in Table-I. In the present work we have made choice of the values for
different parameters to study the physical behaviour for a stable stellar model of gravastar. Again, we also
find different constants for a particular choice of $\xi = 0.24 $ by varying the mass value of gravastar
as shown in Table-II. For a given combination of $M$ and $R_{2}$ it would provide an unique solution,
but we have chosen the values to satisfy the ratio $\frac{2M}{r}< \frac{8}{9}$ and $Z_{s} < 2 $ for a stable
model. As long as this relation holds good we can choose the set of $M$ and $R_{2}$  which provide similar
behaviour as obtained in the paper.

\begin{table}[thbp]\label{tbl1}
\begin{center}
\tabcolsep=0.11cm
\begin{tabular}{|c| c| c| c| c| c| c | } \hline

$\xi$ & $\Lambda$ & $C_{1}$ & $C_{2}$ & $C_{3}$ & $C_{4}$ & Predicted $R_{2}$ (km.)\\ \hline
0.228 & 0.002902 & -0.02501 & 0.9550  & 1.8607 & 0.00005636 & 10.055 \\ \hline
0.23 & 0.002268 & -0.02436 & 0.9589  & 1.9378 & 0.00009521 & 10.061 \\  \hline
0.235 & -0.000056 & -0.02197& 0.9694  & 2.1556 & 0.00003331 & 10.070 \\  \hline
0.24 & -0.0047073  & -0.01721 & 0.9793  & 2.4525 & 0.00115929 & 10.087 \\  \hline
\end{tabular}
\caption{Different values of constants for gravastar having $M= 3.75 M_{\odot}$ and $R_{1} = 10$ km.}
\end{center}
\end{table}

\begin{table}[thbp]\label{Table2}
\begin{center}
\tabcolsep=0.11cm
\begin{tabular}{|c| c| c| c| c| c|} \hline

M & $\Lambda$ & $C_{1}$ & $C_{2}$ & $C_{3}$ & $C_{4}$ \\ \hline
3.5 & 0.004137 & -0.02494 & 0.97375 & 2.1948 & 0.0004616 \\ \hline
3.75 & -0.005467  & -0.01673 & 0.99017  & 2.4684 & 0.0012311 \\  \hline
4.0 & -0.015072 & -0.00852 & 0.99387 & 2.7420 & 0.0020006 \\  \hline
4.25 & -0.024678  & -0.00324 & 0.99582  & 3.0154 & 0.0027701 \\  \hline
\end{tabular}
\caption{Different values of constants for gravastar keeping $\xi= 0.24$, $R_{1} = 10$ km and $R_{2} = 10.05$ km.}
\end{center}
\end{table}

\section{Physical features of the model}\label{sec5}

The intermediate thin shell is assumed to be very thin but having matter of extremely high density, i.e.,
$R_{2}-R_{1}= \epsilon \ll 1 $ where $ r\ (= R_{1} = R)$ is the interior radius of the gravastar
and $R_{2} (= R + \epsilon)$ is the external radius. With these assumptions we have studied the
nature of various physical features of the model, $viz.,$ the proper thickness (or length), energy, entropy, and surface
redshift of the shell. Thus we obtain a set of exact solutions within the shell.

\subsection{Pressure and Matter density}

Inserting the prescribed EOS within the shell, i.e., $ p= \rho$ in Eq.~(\ref{eq11}) along with metric function
of Eq.~(\ref{eq23}) for the thin shell, we obtain the pressure as well as the matter density of the shell as
\begin{equation}\label{eq31}
p = \rho =\rho'_{0}{r}^{{\frac{4(3\xi-1)(4\xi-1)}{(6\xi-1 )^2}}},
\end{equation}
where $\rho'_{0} = \rho_{0} e^{-{\frac { ( 4\xi-1 ) C_4}{(6\xi-1)^2}}}$ and $\rho_{0}$ is an integration constant. The variation of the matter density as well as pressure over the shell is shown in Fig. 1 which shows that the matter density is increasing from the interior
boundary to the exterior boundary.
\begin{figure}[ht]
\label{density}
\centering
\includegraphics[width=8.41cm,height= 5.5 cm]{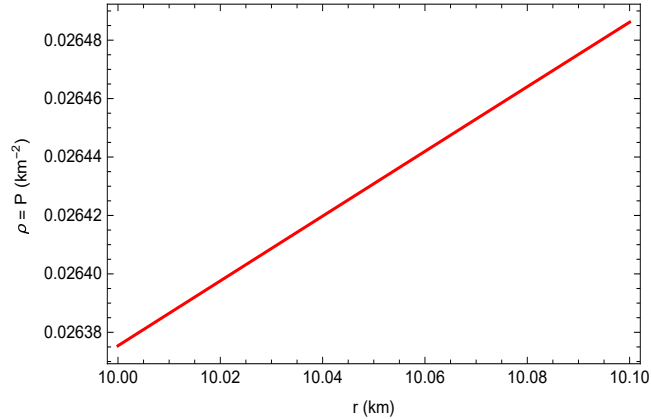}
\caption{Variation of the pressure or the matter density $\rho = p$ ($km^{-2}$) of the shell with respect to r for $M=3.75M_{\odot}$ (taking $\xi = 0.235$, $\rho_{0}= 0.01$)}.
\end{figure}

\subsection{Proper length}
According to the conjecture of Mazur and Mottola, the stiff fluid of the shell is situated between the
junction of two spacetimes. The length span of the shell is from $R_{1} = R$ ($\it i.e.,$ the phase
boundary between the interior and the shell) to $R_{2} (= R + \epsilon)$ ($\it i.e.,$ the phase boundary
between the shell and the exterior spacetime), where $\epsilon << 1$. So, one can calculate the proper
thickness between these two interfaces and the proper length or the proper thickness of the shell can be
determined using the following formula:

\begin{equation}\label{eq32}
   \ell = \int _{R}^{R+\epsilon} \sqrt{e^{\mu} } dr = \int _{R}^{R+\epsilon} \frac{dr}{ \sqrt{-12\xi \ln(r)+2 \ln(r) + C_{3}} }.
\end{equation}

To solve the above equation, let us consider that $\frac{df(r)}{dr} =\frac{1}{\sqrt{e^{-\mu}}} $, so that we get
\begin{equation}\label{eq33}
\ell= \int _{R}^{R+\epsilon} \frac{df(r)}{dr} dr = f(R+ \epsilon)- f(R)=f(R)+ \epsilon f'(R)+\frac{\epsilon^2}{2!} f''(R)- f(R)=\epsilon f'(R)+\frac{\epsilon^2}{2!} f''(R).
\end{equation}

In order to calculate the thickness of thin shell we have adopted the thin shell approximation and restricted
upto second order term of thickness parameter $i.e.,$ the order of $\epsilon^2$ and eventually we have obtained
the proper thickness of the shell as
\begin{eqnarray}\label{eq34}
\ell = \int _{R}^{R+\epsilon} \sqrt{e^{\mu} } dr= \frac{1}{2}\frac{-24\epsilon (\xi-\frac{1}{6}) R \ln(R)+{\epsilon}^{2}(6\xi-1)+2R\epsilon C_3}{ \left((-12\xi+2 )\ln(R)+C_3 \right)^{3/2}R}.
\end{eqnarray}

Variation of proper length within the thin shell has been shown in Fig.~2. The plot shows that the
proper length increases monotonically with the increase of the thickness of the shell.

\begin{figure}[ht] \label{PL}
\centering
\includegraphics[width=8.41cm,height= 5.5 cm]{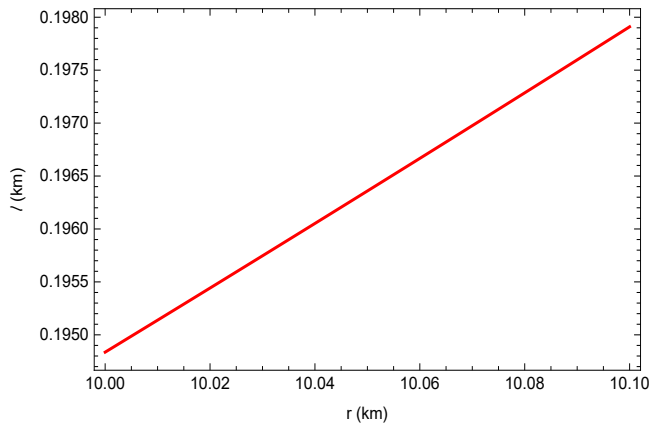}
\caption{Proper length $\ell$(km) of the shell is plotted with respect to the radial distance of the shell $r$ (km) for $M=3.75M_{\odot}$ (taking $\xi = 0.235$).}
\end{figure}

\subsection{Energy}
In the interior region as we consider the EOS, $p = - \rho$, one can argue that this indicates the negative
energy region confirming the repulsive nature of the core. However, the energy within the shell and can be found out to be
The energy $E$ within the shell can be calculated as
\begin{equation}\label{eq35}
{\it E}=\int _{R}^{R+\epsilon} 4\pi r^{2} \rho dr = {\frac{4(6\xi-1)^{2}\pi \rho_0}{156 \xi^{2}-64 \xi+7} \left[ r^{{\frac {156 \xi^2-64\xi+7}{(6\xi-1)^2}}} \right]_{R}^{R+\epsilon}}
\end{equation}
Variation of the energy over the
thin shell has been shown in Fig.~3. This figure demonstrates that the energy is increasing with the increment
of thickness of the shell.

\begin{figure}[ht] \label{energy}
\centering
\includegraphics[width=8.41cm,height= 5.5 cm]{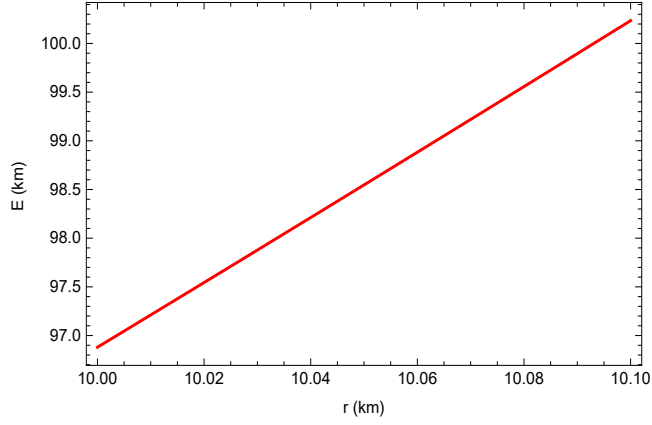}
\caption{Energy E (km) within the shell is plotted with respect to the radial distance of the shell $r$ (km) for $M=3.75M_{\odot}$ (taking $\xi = 0.235$,  $\rho_{0}= 0.01$).}
\end{figure}

\subsection{Entropy}
Following Mazur and Mottola prescription~\cite{Mazur2001}, we can determine the entropy of the thin shell using the following equation,
\begin{equation}\label{eq36}
    S= \int_{R}^{R+\epsilon} 4\pi r^{2} s(r) \sqrt{e^{\mu(r)}} dr
\end{equation}
where $s(r)$ is the entropy density for local temperature $T(r)$ and can be
written as
\begin{equation}\label{eq37}
s(r) = \frac{\alpha^2k_B^2T(r)}{4\pi\hbar^2 } =\alpha\left(\frac{k_B}{\hbar}\right)\sqrt{\frac{p}{2 \pi}}
\end{equation}
where $\alpha$ is a dimensionless constant. Here we have assumed in the Planckian units, i.e.,
$k_{B}$ = $\hbar$ = 1 along with the geometrized units, i.e., $ G = c = 1$ as mentioned earlier.
In order to calculate the entropy of system we have adopted the thin shell approximation and
restricted upto second order term of thickness parameter $i.e.$ order of $\epsilon^2$ and the
entropy of the fluid within the shell can be given by,
\begin{eqnarray}\label{eq38}
S(R) &=&\frac{ \sqrt{\pi}\alpha k_B\sqrt{2}}{h}\left[\epsilon \sqrt{{\frac{\rho_0{R}^{4{\frac{(4\xi-1)(3\xi-1)}{(6\xi-1)^{2}}}}}{-12\xi \ln(R) +2\ln(R) +C_3}}}+\frac{\epsilon^2}{4} \left( {\frac{4\rho_0 (4\xi-1)(3\xi-1)R^{{\frac{4(4\xi-1)(3\xi-1)}{(6\xi-1)^2}}}}{(6\xi-1)^{2}r (-12\xi \ln(R)+2\ln(R) +C_3)}} \right.\right.\nonumber\\
&&
\left.\left.-{\frac {\rho_0 {R}^{4{\frac { (4\xi-1)  \left( 3\xi-1 \right) }{ (6\xi-1)^2}}}}{ \left(-12\xi \ln(R) +2\ln(R) +C_{3} \right)^{2}} \left(-{\frac{12 \xi}{R}}+\frac{2}{R} \right) }\right)
{\frac {1}{\sqrt {{\frac {\rho_0 {R}^{{\frac{4(4\xi-1)  \left( 3
\xi-1 \right)}{(6\xi-1)^2}}} }{-12\xi \ln(R) +2\ln(R) +C_3}}}}}\right].
\end{eqnarray}

Following Ref.~\cite{Usmani2011} it can be shown that (i) the thickness of the shell is negligibly small
compared to its position from the center of the gravastar ($i.e.$, if $\epsilon << R$), and (ii) the entropy
depends on the thickness of the shell. Variation of entropy within the shell is shown in Fig.~4.

\begin{figure}[ht] \label{entropy}
\centering
\includegraphics[width=8.41cm,height= 5.5 cm]{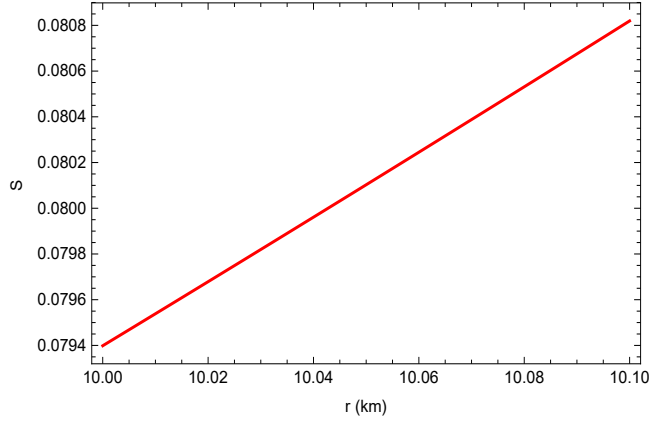}
\caption{Variation of the entropy (S)  of the shell with respect to $r$ for $M=3.75M_{\odot}$ (taking $\xi = 0.235$,  $\rho_{0}= 0.01$).}
\end{figure}

\section{Junction Condition}\label{sec6}
We know that gravastar consists of three regions, viz., interior (I), exterior (III) and intermediate shell (II).
This shell maintains the connection between interior region ($I$) and exterior region ($II$). So it plays
a very crucial part in the configuration of gravastar. This makes a geodetically complete manifold with a
matter shell at the surface $r = R$. Thus a single manifold characterizes the gravastar configuration. At
the junction there has to be a smooth matching between regions I and III, according to the fundamental
junction condition. Though the metric coefficients are continuous at the junction surface but the derivatives
of these metric coefficients may not be continuous there. Now we will use Darmois-Israel condition~\cite{Darmois1927,Israel1966,Israel1967}
to compute the surface stresses at the junction interface. The intrinsic surface stress energy tensor
$S_{ij}$ is given by the Lanczos equation \cite{Lanczos1924,Sen1924,Perry1992,Musgrave1996} in the following form:

\begin{equation}\label{eq41}
    S_{j}^{i}= - \frac{1}{8\pi}\left[K_{j}^{i}- \delta_{j}^{i} K_{k}^{k}\right].
\end{equation}

The discontinuity in the second fundamental form is given by
\begin{equation}\label{eq42}
    K_{ij}= K_{ij}^{+} -  K_{ij}^{-}.
\end{equation}

Here the second elemental form can be written as
\begin{equation}\label{eq43}
K_{ij}^{\pm} = - \eta_{\nu}^{\pm}\left[\frac{\partial^{2} X_{\nu}}{\partial \zeta^{i} \partial \zeta^{j}} + \Gamma^{\nu}_{\alpha\beta}\frac{\partial X^{\alpha}}{\partial \zeta^{i}} \frac{\partial X^{\beta}}{\partial \zeta^{j}} \right]  \vert_{S}
\end{equation}
with the unit normal vector $\eta_{\nu}^{\pm} = \pm \left|g^{\alpha \beta} \frac{\partial f}{\partial X^{\alpha}}
\frac{\partial f}{\partial X^{\beta}} \right|^{\frac{1}{2}} \frac{\partial f}{\partial X^{\nu}}$ with $\eta^{\nu}\eta_{\nu} = 1$.
Here $\zeta^{i}$ is the intrinsic coordinate on the shell and $f(X^{\alpha}(\zeta^{i})) = 0$ is the parametric
equation of the shell. Here $+$ and $-$ correspond to Schwarzschild spacetime and the interior spacetime of the
gravastar respectively.

Considering spherically symmetric spacetime following Lanczos equation, the surface stress energy tensor
can be written as $S_{ij}= diag (-\Sigma,P, P, P)$, where $\Sigma$ and $P$ are surface energy density and
surface pressure respectively. The parameters $\Sigma$ and $P$ can be expressed by the following equations as
\begin{equation}\label{eq44}
   \Sigma = \left[ - \frac{1}{4\pi R} \sqrt{f} \right]_{-}^{+},
\end{equation}
\begin{equation}\label{eq45}
    P = \left[ - \frac{\Sigma}{2} + \frac{f'}{16 \pi \sqrt{f}} \right]_{-}^{+}.
\end{equation}
So, using the above two Eqs~(\ref{eq44}) and (\ref{eq45}) we have obtained the surface energy density and surface pressure respectively as,
\begin{eqnarray}\label{eq46}
\Sigma={\frac {1}{12\pi R} \left( \sqrt{3R^2C_1+9}-\sqrt{9-{\frac{18M}{R}}-3\Lambda R^2} \right)},
\end{eqnarray}

\begin{eqnarray}\label{eq47}
\textit{P}={\frac {3}{8\pi R} \left({{\frac{\left(1-{\frac {M}{R}}-\frac{2 \Lambda R^2}{3} \right)}{\sqrt {9-{\frac{18 M}{R}}-3\Lambda R^2}}}}-{\frac {\frac{2}{3}{R}^{2}C_{{1}}+1}{\sqrt{3R^2C_1+9}}} \right) }.
\end{eqnarray}

Variation of the surface energy density is shown in Fig.~6. We have observed that both the parameters
remain positive throughout the shell which immediately indicate that null energy condition has been
satisfied for the formation of thin shell. Here, in Fig.~6, we have showed the variation of the surface
energy density ($\Sigma$) for four different $\xi$ values taking mass of the gravastar  $M=3.75M_{\odot}$
and $R_{1}=10.0$ km and thus we can predict the boundary of the intermediate shell ($R_{2}$) where
the surface energy density ($\Sigma$) goes to zero for different $\xi$ values, which is first observed in our study.
It is also evident that although surface energy density is more for lower $\xi$, it attains vanishing
value earlier (say $\xi=0.28\; (Brown)$ in Fig.~6.

In Table-I, we have seen that the value of $R_{2}$ increases $i.e.,$ the thickness of the boundary
increases for higher $\xi$ values of a particular star. There is a discontinuity of the second
fundamental form at the junction between the two spacetimes which further implies that there must
be a matter component (ultra relativistic fluid) obeying EOS $p= \rho$. This non-interacting matter
or the fluid characterizes the existence of the thin shell of the gravastar.

\begin{figure}[ht] \label{SurED}
\centering
\includegraphics[width=8.41cm,height= 5.5 cm]{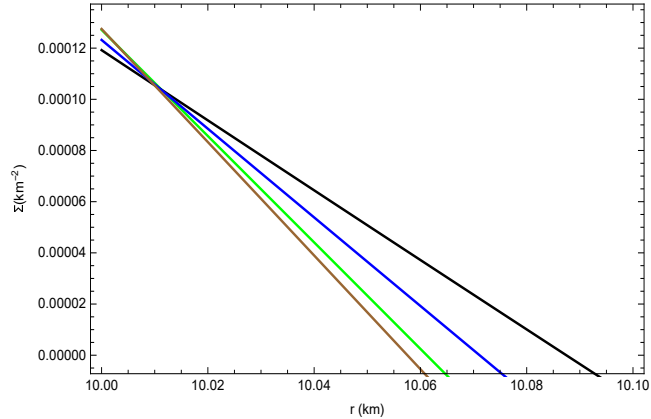}
\caption{Variation of the surface energy density ($\Sigma$) ($km^{-2}$) of the shell with $r$ (km) taking $M=3.75M_{\odot}$ for different $\xi= 0.228 (Brown)$, $\xi= 0.23 (Green)$, $\xi= 0.235 (Blue)$ and  $\xi= 0.24 (Black)$ values}
\end{figure}

Now, it is easy to find the surface mass $m_s$ of the thin shell from the equation of the surface energy density
\begin{eqnarray} \label{eq48}
m_s= 4 \pi r^{2} \Sigma = \frac{R}{3} \left(\sqrt{3R^2 C_1+9}-\sqrt {9-{\frac{18 M}{R}}-3\Lambda R^2} \right), \label{eq36}
\end{eqnarray}
For negative $C_1$, we get a upper limit of radius
$R < \sqrt{ \frac{3}{|C_1|}}$  for which $m_s$ is realistic. Finally we note a limiting value on the radius
\[
 \left( \frac{6 M}{\Lambda -|C_1|}  \right)^{\frac{1}{3}}  <  R < \sqrt{ \frac{3}{|C_1|}}
 \]
with $|C_1| < \Lambda$. The mass $M$, $\Lambda$ and $C_1$ determines the range of values of radius. It is
interesting to note that the mass of a gravastar determines the lower limit of its size.
With the help of the above equation we calculate the total mass of the gravastar in terms of the mass of the thin shell ($m_s$) as
\begin{eqnarray}\label{eq49}
M=\frac{2R m_s \sqrt{3 R^2 C_1+9} -R^4(\Lambda + C_1)-3{m_s}^2}{6R}.
\end{eqnarray}

\section{Stability}\label{sec7}
In this section, we have studied the stability of our theoretical model by two methods: (i) Surface Redshift and (ii) Entropy maximization. The methods are described below:
\subsection{Surface Redshift}
Study of surface redshift of the gravastars can be considered as one of the most important source of information
regarding it’s stability and detection. The surface gravitational redshift can be defined as
$Z_{s} = \frac{\Delta \lambda}{\lambda_{e}} = \frac{\lambda_{0}}{\lambda_{e}}$, where $\Delta$
represents the fractional change in wavelength between the emitted signal $\lambda_{e}$ and observed signal $\lambda_{0}$.
Buchdahl claimed that for an isotropic, static, perfect fluid distribution the value of surface
redshift should not exceed $2$, $i.e.$, $Z_{s} < 2 $~\cite{Buchdahl1959,Straumann1984,Bohmer2006}. Ivanov~\cite{Ivanov2002} claimed for anisotropic fluid distribution
it can increase upto $3.84$. Barraco and Hamity~\cite{Barraco2002} showed that $Z_{s} \leq 2 $ for an isotropic fluid distribution when
the cosmological constant is absent. But Bohmer and Harko~\cite{Bohmer2006} proved that in presence of
cosmological constant of an anisotropic star $Z_{s} \leq 5 $. To calculate the surface redshift we have
used the following equation

\begin{equation}\label{eq39}
	Z_{s}= -1+ \frac{1}{\sqrt{g_{tt}}}
\end{equation}
eventually we get the surface redshift as,
\begin{equation}\label{eq40}
	Z_{s}=  -1+ \frac{1}{C_{4}^{\frac{1}{2}}\;  r^{\left(\frac{2- 6 \xi}{6 \xi- 1}\right)}}
\end{equation}

We have plotted the variation of surface redshift in Fig.~6. From the figure it can be observed
that the value of $Z_{s}$ lies within 1 throughout the thin shell. So, our present model of gravastar can be
claimed to be stable as well as physically acceptable.

\begin{figure}[ht] \label{SurRed}
	\centering
	\includegraphics[width=8.41cm,height= 5.5 cm]{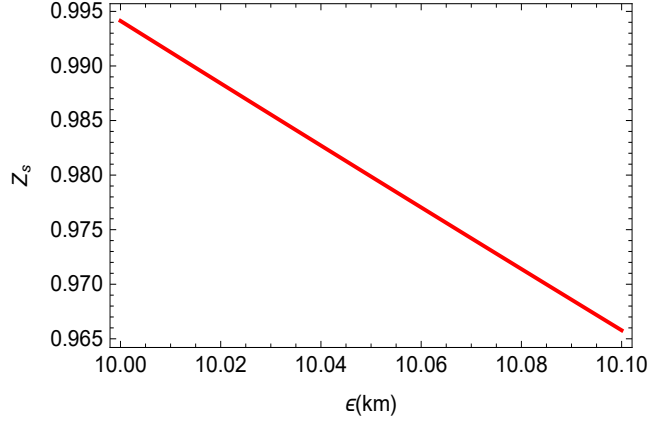}
	\caption{Variation of surface redshift of the shell with its thickness  $\epsilon$ (km) for $M=3.75M_{\odot}$ (taking $\xi = 0.235$).}
\end{figure}

\subsection{Entropy maximization}
Now to check the stability of the present study on gravastar in Rastall gravity we have followed the
technique proposed by Mazur and Mottola~\cite{Mazur2001,Mazur2004}. In order to accomplish
this we have applied the technique of maximization of the entropy functional with respect to the
mass function $M(r)$ and checked the sign of second variation of the entropy
function. The first variation must vanish at the boundaries of the intermediate thin shell of the gravastar, which are fixed at $r=R_1$ and $r=R_2$ respectively  i.e., $\delta S=0$

Now, the entropy functional can be written as
\begin{equation}
S = \frac{\sqrt{2}\alpha k_B}{2 \hbar }\int^{R_2}_{R_1} r dr \left(\frac{dM}{dr}\right)^{\frac{1}{2}} \frac{1}{\sqrt{g(r)}}, \label{eq41}
\end{equation}
here $g(r)=1-\frac{2M}{r}-\frac{\Lambda r^2}{3}$.

Now from Eqs. (\ref{eq22}), (\ref{eq25}) and (\ref{eq26}) one can find the form of $M(r)$ in Rastall theory of gravity as
\begin{equation}
M(r)=-r \left(( 6\xi-1) \ln \left(\frac{R_1}{r} \right) +\frac{\Lambda {r}^{2}}{6}+\frac{C_1{R_1}^2}{6}\right). \label{eq42}
\end{equation}

Hence, we can obtain the second variation of the entropy function as
\begin{eqnarray}\label{eq43}
\delta^2 S=\frac{\sqrt{2}\alpha k_B}{2 \hbar }\int^{R_2}_{R_1} rdr \left(\frac{dM}{dr}\right)^{\frac{-3}{2}} \frac{1}{\sqrt{g(r)}}\left[-\left(\frac{d(\delta M)}{dr}\right)^2+\frac{(\delta M)^2}{g^2r^{2}}\left(\frac{dM}{dr}\right)\left(1+\frac{dM}{dr}\right)\right].
\end{eqnarray}

Here we have considered the linear combination of $M(r)$ as $\delta M = \beta \psi$, where $\psi$ vanishes at the boundaries.
Now inserting this into Eq. \ref{eq43}, and integrating by parts keeping $\delta M =0$ at the extreme points of the thin shell, eventually we left with
\begin{equation}
\delta^2 S = -\frac{\sqrt{2}\alpha k_B}{2 \hbar }\int^{R_2}_{R_1} r dr \left(\frac{dM}{dr}\right)^{\frac{-3}{2}} \frac{1}{\sqrt{g(r)}}\beta^2\left[\frac{d\psi}{dr}\right]^2 < 0. \label{eq44}
\end{equation}

It clearly shows that, the entropy function in Rastal gravity  attains the maximum value for all the radial variations which are vanished at the end points of the boundary of the shell of the gravastar. So we can conclude that the perturbation in the fluid of the intermediate shell region of the gravastar causes a decrease in entropy in the region II which is further indicating that our solutions are stable against small perturbations with fixed end points. In essence, the effect of modified gravity does not affect the stability of the gravastar.

\section{Discussions and Conclusions}\label{sec8}
In the paper  a unique stellar model of gravastar is studied following the conjecture of Mazur-Mottola in
the Rastall's theory of gravity. We obtain a set of exact,
singularity free and physically acceptable solutions with a number of interesting and satisfactory  properties of the gravastar in
Rastall gravity. The following salient features of  gravastar are noted :\\

(1) $\textbf{Interior region}$ : Using the EOS given by Eq.~(\ref{eq12}) and the conservation
equation Eq.~(\ref{eq11}) it is found that the matter density as well as the pressure remains
constant in the interior. Again using field Eqs.~(\ref{eq8})- (\ref{eq10}),  the metric
functions $e^{\mu(r)}$ and $e^{\nu(r)}$ are determined. It is evident  that both the functions
are continuous at the origin $r = 0$, and thus free from  central singularity. We have also determined the active gravitational mass of the interior.\\

(2) $\textbf{Intermediate thin shell}$ : In the intermediate thin shell region, one can apply the thin
shell approximation and compute the corresponding metric functions.  Both the metric functions remain finite and positive over the shell. The metric functions are  modified due to the Rastall gravity effects which  depend on the Rastall parameter $\xi$.\\

(3) $\textbf{Physical parameters of the Shell}$: Various physical parameters associated with the shell
are  computed which are modified in the  Rastall gravity. The
detail of the parameters are as follows:\\

(i) \textbf{Pressure and matter density}: The matter density or the pressure of the shell
is plotted against the radial distance $r$  in Fig.~1. The variation of matter density or pressure for the shell
is found to be positive which  is monotonically  increasing towards the  exterior surface. This
demands that the shell becomes more denser at the exterior boundary than  the interior boundary.\\

(ii) $\textbf{Proper thickness}$: The proper thickness of the shell is determined adopting the
thin shell approximation  which is    physically acceptable. The proper thickness or
proper length $\ell$ of the shell is found to be monotonically increasing in nature from the interior junction
to the exterior junction which is shown in Fig.~2.\\

(iii) $\textbf{Energy}$: The energy of the shell has been obtained in Eq.~(\ref{eq35}) and the variation with respect to
radial parameter is shown in Fig.~3. The variation of energy is similar to that of the matter density. It   satisfies
the requirement that the energy of the shell increases with the increase in the radial distance.\\

(iv) $\textbf{Entropy}$: The entropy $S$ within the shell is given by  Eq.~(\ref{eq38}) and the variation of entropy is shown in Fig.~4.  The entropy is gradually increasing with respect to the radial distance $r$ indicating a maximum value of entropy on the surface of the gravastar which satisfies the physical validity condition.
In order to calculate the entropy of the thin shell we assumed thin shell approximation and restricted upto
second order term of thickness parameter $\epsilon$ ($i.e.,$ the order of $\epsilon^2$).\\

(4) $\textbf{Junction condition}$: The junction condition for the formation of thin shell
between the interior and exterior spacetimes is considered. Following the condition of Darmois and Israel we  study
the variation of the surface energy density due to the formation of thin shell, which is plotted in Fig.~5
for different $\xi$ values. Both the parameters remain positive which indicates that the thin shell
satisfies the weak and dominant energy conditions.  In Fig.~5, we plot the variation of the surface energy density ($\Sigma$) for four
different $\xi$ values taking mass of the gravastar $M=3.75M_{\odot}$ and $R_{1}=10.0$ km. This determines the variation of boundary of the gravastar with the Rastall parameter. \\

(5) $\textbf{Stability}$: The stability of the model is studied considering the surface redshift analysis and the entropy maximization technique. It is  found that our model is stable and physically acceptable. It is found that : \\

(i) $\textbf{Surface redshift}$:The stability of the gravastar is checked using  surface redshift analysis.
For any stable stellar model the value of surface redshift lies within $2$. In the present case we have found that our model is
stable under surface redshift analysis as can be noted from Fig.~6.\\

(ii) $\textbf{Entropy Maximization}$: We have checked the stability of  gravastar models in Rastall
gravity following the technique proposed by Mazur and Mottola~\cite{Mazur2001,Mazur2004}.We employ the
 maximization of the entropy with respect to the mass function $M(r)$ and checked
the sign of second variation of the entropy  for its consistency. Eq. (\ref{eq44}) shows that the entropy function in
Rastall gravity  attains the maximum value for all the radial variations which are vanished at the end points
of the boundary of the shell of the gravastar,  perturbation in the fluid of the intermediate shell region of
the gravastar causes a decrease in entropy in the region II which is further indicating that our solutions are
stable against small perturbations with fixed end points.\\

(6) We have obtained the five different constants for different values of $\xi$ values which is
shown in Table-I by choosing the mass of the gravastar $M= 3.75 M_{\odot}$ and interior radius
$R_{1} = 10$ km. In Table-II, we have also predicted the variation of the exterior boundary of
the thin of shell ($R_{2}$) of the gravastar with the Rastall parameter ($\xi$). Thus we have examined the
effect of Rastall gravity ($\xi$) in gravastar which is a new result.\\

(7) Again, we also find different metric parameters for a particular choice of $\xi = 0.24 $ by varying the
mass value of gravastar which is shown in Table-II. For a given combination of $M$ and $R_{2}$ it
would provide an unique solution, but we have chosen the values to satisfy the ratio $\frac{2M}{r}< \frac{8}{9}$
and $Z_{s} < 2 $ for a stable model.\\

From all the above analysis one can easily conclude that the gravastar may exist under
the framework of Rastall's theory of gravity. Unlike the previous works on gravastars, we
have taken the thin shell approximation up to the second-order which provides more accurate
results of the physical parameters of the shell. It is interesting to note that under the framework of of Rastall gravity the form of different physical parameters viz., energy, entropy, proper length etc. have been modified than those obtained in Einstein's GR or other alternative gravity theories. One can get back the expressions of these parameters as obatained in Einstein's GR by putting $\xi=0$ and restricting to the first-order approximation
of the thickness parameter ($\epsilon$) of the shell. Also, unlike
Einstein's GR one can observe that
the effect of Rastall gravity demands the exterior region of the gravastar to be de-Sitter type which is a clear indication of the existence of
dark energy at the exterior region and its effect on the formation of stable gravastar under the framework of Rastall gravity.

\subsection{Possibility of detection of gravastar:} Till date there are no direct evidences in favour of gravastar but
few indirect evidences are available in the literature for predicting the  existence and future detection of gravastar  \cite{Abbott2016,Sakai2014,Kubo2016,Cardoso1,Cardoso2,Chirenti2016,Uchikata2016}. The concept for possible detection mechanism
of gravastar was first proposed by Sakai et al.~\cite{Kubo2016} through the study of gravastar shadows. Another
possible method for the detection of gravastar may be employed by gravitational lensing method as suggested by
Kubo and Sakai~\cite{Kubo2016} where they have claimed that  in a gravastar microlensing effects of larger
maximal luminosity compared to black holes of the same mass might occur. According to Cardoso et al.~\cite{Cardoso1,Cardoso2}
the ringdown signal of GW 150914~\cite{Abbott2016} detected by interferometric LIGO detectors are supposed to be
generated by objects without event horizon which might be a due to gravastar, though it is yet to be confirmed. Very recently
in analysing the image obtained in the First M87 Event Horizon Telescope (EHT) result \cite {Akiyama2019} it is
not ruled out the possibility that  the produced shadow may be due to gravastar.
Shadow can be produced by
any compact object having a spacetime which is characterized by unstable circular photon orbits as shown
by Mizuno et al. \cite{Mizuno2018}.  According to Kubo and Sakai \cite{Kubo2016} gravastars can possess unstable
circular photon orbits. In future, if similar effects mentioned above are detected observationally then it will be a good platform for comparing GR with  Rastall gravity. \\

Analyzing the above points for our present study, we can conclude that, we obtain  solutions in the Rastall theory of
gravity which describe gravastars satisfactorily. For a gravastar in Rastall gravity an upper limit of
radius is obtained determined by an integration constant $|C_1| < \Lambda$. We note a new result here that
the mass of a gravastar determines its size. We have obtained a set of physically acceptable and non singular
solution of the gravastar which immediately overcome the  singularity problem and existence
of event horizon of black hole. Analyzing we claim that our model of gravastar is stable under Rastall theory 
of gravity which is conceptually different from Einstein's GR. 

As a final comment we can say that, though there is no direct observational evidences are available till now which 
can differentiate a black hole from a gravastar. Again the recent observations of GW190521 clearly pointing out that 
the black hole theory is inconsistent with the observational results. Thus it can be argued that the possible black hole 
might be a possible gravastar.

\section{Acknowledgements}
SG and AD is thankful to the Department of Physics, IIEST Shibpur for providing research facilities
and also thankful to Dr. Saibal Ray of G.C.E.C.T, Kolkata and Prof. B.K. Guha of Dept. of Physics,
IIEST, Shibpur for their help and support. SG is also thankful to the Directorate of Legal Metrology
under the Department of Consumer Affairs, West Bengal for their support. SD is thankful to UGC, New
Delhi for financial support. The authors would like to thank IUCAA Centre for Astronomy Research and Development (ICARD),
NBU for extending research facilities. AC would like to thank University of North Bengal for awarding Senior
Research Fellowship. BCP would like to thank DST-SERB Govt. of India (File No.: EMR/2016/005734) for a project.

\end{document}